\documentclass[aps,pra,superscriptaddress,reprint,
dvipsnames]{revtex4-1}
\usepackage{amsmath,amssymb,bm}
\usepackage{graphicx}
\usepackage{float}
\usepackage{braket}
\usepackage{dsfont}
\usepackage{appendix}
\usepackage{youngtab}
\usepackage{ytableau}
\usepackage{multirow}
\usepackage{physics}
\usepackage{todonotes}
\usepackage[utf8]{inputenc}
\usepackage[T1]{fontenc}
\usepackage{appendix}
\usepackage[caption=false,,subrefformat=parens,labelformat=parens]{subfig}
\usepackage{soul}
\setstcolor{blue}
\newcommand{\yongtwo}{\Yvcentermath1\Yboxdim{5pt}\yng(2)}
\newcommand{\yongoneone}{\Yvcentermath1\Yboxdim{5pt}\yng(1,1)}
\newcommand{\yongtwoone}{\Yvcentermath1\Yboxdim{5pt}\yng(2,1)}
\newcommand{\yongthree}{\Yvcentermath1\Yboxdim{5pt}\yng(3)}
\newcommand{\yongoneoneone}{\Yvcentermath1\Yboxdim{5pt}\yng(1,1,1)}
\newcommand{\yongthreeone}{\Yvcentermath1\Yboxdim{5pt}\yng(3,1)}
\newcommand{\yongtwotwo}{\Yvcentermath1\Yboxdim{5pt}\yng(2,2)}
\newcommand{\yongtwooneone}{\Yvcentermath1\Yboxdim{5pt}\yng(2,1,1)}
\newcommand{\yongfour}{\Yvcentermath1\Yboxdim{5pt}\yng(4)}
\newcommand{\yongfourones}{\Yvcentermath1\Yboxdim{5pt}\yng(1,1,1,1)}

\newcommand{\yongwhite}{{\scalebox{.4}{\ytableausetup{nosmalltableaux}
\begin{ytableau} *(white)\end{ytableau}}}}
\newcommand{\yongblue}{{\scalebox{.4}{\ytableausetup{nosmalltableaux}
\begin{ytableau} *(blue)\end{ytableau}}}}
\newcommand{\yongCyan}{{\scalebox{.4}{\ytableausetup{nosmalltableaux}
\begin{ytableau} *(Cyan)\end{ytableau}}}}
\newcommand{\yonggreen}{{\scalebox{.4}{\ytableausetup{nosmalltableaux}
\begin{ytableau} *(green)\end{ytableau}}}}
\newcommand{\yongForestGreen}{{\scalebox{.4}{\ytableausetup{nosmalltableaux}
\begin{ytableau} *(ForestGreen)\end{ytableau}}}}
\newcommand{\yongPurple}{{\scalebox{.4}{\ytableausetup{nosmalltableaux}
\begin{ytableau} *(Purple)\end{ytableau}}}}
\newcommand{\yongLavender}{{\scalebox{.4}{\ytableausetup{nosmalltableaux}
\begin{ytableau} *(Lavender)\end{ytableau}}}}
\newcommand{\yongorange}{{\scalebox{.4}{\ytableausetup{nosmalltableaux}
\begin{ytableau} *(orange)\end{ytableau}}}}
\newcommand{\yongOrchid}{{\scalebox{.4}{\ytableausetup{nosmalltableaux}
\begin{ytableau} *(Orchid)\end{ytableau}}}}
\newcommand{\yongDandelion}{{\scalebox{.4}{\ytableausetup{nosmalltableaux}
\begin{ytableau} *(Dandelion)\end{ytableau}}}}
\newcommand{\yongyellow}{{\scalebox{.4}{\ytableausetup{nosmalltableaux}
\begin{ytableau} *(yellow)\end{ytableau}}}}
\newcommand{\yongBrickRed}{{\scalebox{.4}{\ytableausetup{nosmalltableaux}
\begin{ytableau} *(BrickRed)\end{ytableau}}}}
\newcommand{\yongred}{{\scalebox{.4}{\ytableausetup{nosmalltableaux}
\begin{ytableau} *(red)\end{ytableau}}}}
\newcommand{\yongblack}{{\scalebox{.4}{\ytableausetup{nosmalltableaux}
\begin{ytableau} *(black)\end{ytableau}}}}

\newcommand{\svdots}{%
  \vbox{
    \small \baselineskip 4pt
    \lineskiplimit -5pt
    \hbox {.}\hbox {.}\hbox {.}
  }%
}

\newcommand{\vcenterytableau}{%
  \begin{gathered}
  \tiny\yng(1,1)\\\baselineskip -5pt\tiny\svdots\\
  \tiny\yng(1)
  \end{gathered}
}

\usepackage{soul}

\begin{document}
\title{Interferometrically estimating
a quadratic form for any immanant of a matrix
and its permutations}
\author{Aeysha Khalique}
\affiliation{%
	School of Natural Sciences, National University of Sciences and Technology,
	H-12 Islamabad, Pakistan%
	}
\affiliation{National Centre for Physics (NCP), Shahdra Valley Road, Islamabad 44000, Pakistan}
\author{Hubert de~Guise}
\email{hubert.deguise@lakeheadu.ca}
\affiliation{Department of Physics, Lakehead University, Thunder Bay, Ontario P7B 5E1, Canada}
\author{Barry C. Sanders}
\email{sandersb@ucalgary.ca}
\affiliation{%
	Institute for Quantum Science and Technology, 
	University of Calgary, Alberta T2N 1N4, Canada}
\affiliation{%
	Program in Quantum Information Science,
	Canadian Institute for Advanced Research, Toronto, Ontario M5G 1Z8, Canada}
\date{\today}
\begin{abstract}
We devise a multiphoton interferometry scheme for 
sampling a quadratic function of a specific immanant
for any
submatrix of a unitary matrix and its row permutations.
The full unitary matrix describes a passive, linear interferometer,
and its submatrix is used when photons enter in and
are detected at subsets of possible input and output channels. 
Immanants are mathematical constructs that interpolate between the permanent and determinant;
contrary to determinants and permanents, which have meaningful physical applications,
immanants are devoid of physical meaning classically but here are shown to be meaningful in a quantum setting.
Our quadratic form of immanants is sampled
by injecting vacuum and single photons
into interferometer input ports such that the photon arrival times are entangled,
in contrast to previous methods that control arrival times without entangling.
Our method works for any number of photons,
and we solve explicitly the quadratic form
for the two-, three- and four-photon cases. 
\end{abstract}
\maketitle
\section{Introduction}
\label{sec:intro}
Whereas matrix determinants and permanents have immediate physical applications in many-body quantum physics~\cite{BF04}
and quantum information~\cite{AA11},
immanants~\cite{L50,Valiant1979}
only recently fully connected with physical concepts 
via quantum interferometry.
Specifically,
$n$ photons, controllably distinguishable through time delays or polarization~\cite{TGdG+13,dGTP+14,TTS+14,WdGS18},
are injected into an 
$m$-channel interferometer,
and $n$-photon coincidences at the interferometric output ports depend on sums of immanants of a square complex matrix
describing the interferometer.
This coincidence probability in fact depends in general on all immanants of the transformation matrix including the determinant and permanent.
Designing input states that are products of partially distinguishable single-photon input states and produce a scattering amplitude proportional to a \emph{single} immanant
has been impossible thus far,
except for the case of perfect indistinguishability where only the permanent of the appropriate scattering submatrix contributes to the amplitude.
Here, we show how time-bin-entangled input states can lead to coincidence rates that are obtained from quadratic sums of a \emph{single} type of immanant;
the immanants in the sum differ because each is evaluated after some rows or columns of the original submatrix are permuted.

Immanants
of a square complex matrix are defined using a weighted sum
\begin{align}
\label{eq:imm}
\text{imm}^\lambda (U)
        :=\sum_{\sigma\in S_n}\chi^\lambda(\sigma)
            \prod_{\imath=1}^n U_{\imath,\sigma(\imath)},
\end{align}
where $ U\in M_n(\mathbb{C}) $
for $M_n$ denoting an $n\times n$ matrix, $\sigma(\imath)$ a permutation of $\imath\in\mathbb{Z}_+$, and  $\chi^\lambda(\sigma)$ the character of $\sigma\in S_n$
in irreducible representation (irrep)~$\lambda$. The permanent and determinant are special cases of immanants corresponding to
\begin{align}
\lambda=
\begin{cases}
\{n\}
=\tiny{ \yng(2)\cdots\tiny\yng(1)}\; \\ \{1^n\}=
   \tiny{\vcenterytableau}:=\tiny{ \yng(2)\cdots\tiny\yng(1)}^\top
   \end{cases}
\end{align}%
for symmetric and antisymmetric representations of $S_n$, the permutation group of $n$ objects, such that
\begin{equation}
    \chi^{\tiny{ \yng(2)\cdots\tiny\yng(1)}}(\sigma)\equiv 1\forall \sigma
    \end{equation}
and
\begin{equation}
    \chi^{\tiny{ \yng(2)\cdots\tiny\yng(1)}^\top}(\sigma)=\pm 1,
\end{equation}
depending on the parity of $\sigma$;
for the other representations of $S_n$ the characters are not necessarily $\pm 1$,
and other immanants interpolate beautifully between the permanent and the determinant.
Although the immanant is now recognized as being an integral part of calculating interferometric coincidence probabilities,
direct meaning arising from realizing probabilities for a single immanant has been lacking until now.

The computational hardness (\#P-Hard)
of calculating the permanent~\cite{Valiant1979}
underpins significant efforts on the \textsc{BosonSampling} problem~\cite{AA11},
both theoretically~\cite{MGD+14}
and experimentally~\cite{SMH+13,TDH+13,TTS+14,BSV+15,ZLL+18},
with ramifications for the pursuit of quantum supremacy~\cite{ClCl18}.
Unlike the permanent or the determinant that picks up at most a sign under permutations of rows or columns,
an immanant does not necessarily transform back to a multiple of itself under such transformations, highlighting that this 
function cannot possibly describe identical fermion or boson states.
Some immanants are also \#P-Hard~\cite{Hartmann85, BB03}
and,
as effective non-classical functions
of the interferometer transition matrix~$R$,
are especially interesting in their own right. 

Now we explain the sum of product of immanants that is approximately solved through interferometry.
For $\{a_{\sigma\sigma'}\}$ a to-be-determined set of complex coefficients, let $\lambda=\{\lambda_1,\ldots,\lambda_p\},
\lambda_j\ge \lambda_{j+1}, p\le n$ be a fixed partition of $n$, and consider
\begin{align}
    F_\lambda(U_n):=\sum_{\sigma\sigma'} a_{\sigma\sigma'} \text{imm}^\lambda ((U_n)_\sigma)
     \left(\text{imm}^\lambda ((U_n)_{\sigma'})\right)^*,
    \,  
   \label{eq:FlambdaU}
\end{align}
where
the arguments of 
$\text{imm}^\lambda$ in the sum $F_\lambda(U_n)$ are matrices $(U_n)_\sigma$ differing from $U_n$ by a permutation
of rows: thus $\hbox{imm}^\lambda((U_n)_\sigma)$ is the $\lambda$-immanant of $(U_n)_\sigma$.
Our goal is to find $a_{\sigma\sigma’}$ so the quadratic form $F_\lambda(U_n)$ can be sampled using a linear optics setup, thereby connecting the immanant with experimental technology. As some immanants are \#P-Hard~\cite{BB03}, this leads to a broad diversification of the possibilities for empirically studying
\#P-Hard problems.

Our paper is organized as follows: In \S\ref{sec:scheme} we devise the experiment to calculate the quadratic function of each immanant of $U_n$ and explain the algorithm, computational task and general form of the estimate.
In \S\ref{sec:n234}, we apply our general formalism to $n\in\{2,3,4\}$ and give the explicit form of the estimate of the quadratic function in each case. Finally, we conclude in \S\ref{sec:conclusions}.
\section{The Scheme}
\label{sec:scheme}
\begin{figure}
    \includegraphics[scale=0.3]{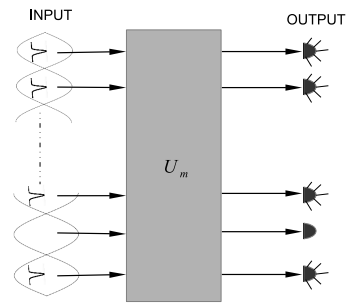}
    \caption{The proposed experiment: The input to the interferometer $U_m$ are the time-bin entangled $n\le m$ single photons with spectral profile and the desired output is the coincidence click at all the detectors at output}
    \label{fig:experiment}
\end{figure}

We devise time-bin entangled input states 
for a passive, lossless, $m$-channel interferometer, with one photon injected into each input port,
such that the output coincidence probabilities sample a quadratic form
$F_\lambda(U_n)$, describing the interferometric transformation of 
an $n\times n$ submatrix~$U_n$ of the $U_m$ matrix. We can assume that $U_m$ is unitary, but in general the submatrix $U_n$ will not be so. The proposed experiment is shown in~Fig.~\ref{fig:experiment}.
The input state is an entangled state of single photon pulses into each input port, with the basis state being the product of single photon Fock states in identical localized wave packets. The entangled time bins are superposition of such product states. Detectors integrate over the whole time-scale and are set up as coincidence detectors.

We now set up the problem as a computational task. Specifically, we discuss the task as a two-party protocol involving the client who provides input and accepts the output as well as the server who computes and produces the output. We also provide the quantum circuit that performs sampling to deliver an approximate solution.

In \S\ref{subsec:computationaltask}, we explain the algorithm and computational task carried out by the proposed experiment. In \S\ref{subsec:estimate}, we give the general form of the estimate of the quadratic function of immanants carried out by the experiment. In \S\ref{subsec:exprealize}, we explain how various components of the proposed scheme can be physically realized.
\subsection{Algorithm and the computational task}
\label{subsec:computationaltask}
Now we explain the algorithm as well as clarify the computational task. The input of the computational task is the value $n\in\mathbb{Z}^+$, $U_n\in M_n(\mathbb{C})$ with~$\mathbb{C}$ the complex number field, the error bound $\epsilon>0$, the values of time delays $\tau\in\mathbb{R}^+$, and the spectral profile $\phi:\mathbb{R}^+\to(0,1]:\omega\mapsto\phi(\omega)$ of each photon. Now we explain the output.

The server executes the procedure by using both a classical computer and a multiphoton, multichannel quantum interferometer.
We begin by explaining the interferometer, which accepts single photon pulses 
\begin{equation}
\ket 1=\int\text{d}\omega\phi(\omega)\ket{1(\omega)}
\label{1photonomega}
\end{equation}
with
\begin{equation}
\ket{1(\omega)}:=a^\dagger (\omega)\ket0 
\label{1photonket}
\end{equation} 
and vacuum $\ket0$ as input with~$a^\dagger(\omega)$ the  creation operator for frequency $\omega$. 

For the multi-channel interferometer
and

\begin{equation}
\bm\omega=(\omega_1\cdots\omega_n),    
\end{equation}
we use the shorthand notation
\begin{equation}
 \ket{\bm1(\bm\omega)}
:=\ket{1(\omega_1)}\cdots\ket{1(\omega_n)}   
\end{equation}
for the products of states
of the type~(\ref{1photonomega}).
Next we explain how to replace this product state by a time-bin entangled state
with time delays
\begin{equation}
\bm\tau
    :=(\tau_1\cdots\tau_n)^\top     
\end{equation}
for the immanant of $U_n$ associated with irrep~$\lambda$ with character $\chi^\lambda(\sigma)$.

For unitary representation
$P(\sigma)$
of~$\sigma\in S_n$,
\begin{equation}
\label{eq:permnotation}
P(\sigma)\ket{\bm1(\bm\omega)}
=\ket{\bm1(\bm\omega_\sigma)},
\end{equation}
where
\begin{equation}
\label{eq:permomega}
    \bm\omega_\sigma
:=\left(\omega_{\sigma(1)}\cdots\omega_{\sigma(n)}\right), 
\end{equation}
for example for $\sigma=(123)$, $\bm\omega_\sigma=(\omega_3,\omega_1,\omega_2)$. We define the input state
\begin{equation}
\label{eq:input1}
\ket{\bm1}^\lambda
  \left(\bm\tau\right)
=\int\frac{\text{d}^n\bm\omega}
  {\sqrt{n!}}
 \phi(\bm\omega)
\text{e}^{-\text{i}\bm\omega\cdot\bm\tau}
\sum_{\sigma\in S_n}
\chi^\lambda(\sigma)
\ket{\bm1(\bm\omega_\sigma)},
\end{equation}
with
\begin{equation}
 \text{d}^n\bm\omega=\text{d}\omega_1\cdots\text{d}\omega_n, \,\phi(\bm\omega):=\phi(\omega_1)\cdots\phi(\omega_n).
\end{equation}
This input state contains as weighting factors the characters of classes for irrep $\lambda$, which ensures that only coefficients of products of one immanant of $U_n$
and its permutation survive in the coincidence probability.

For $n=2$ photons,
and replacing
\begin{equation}
    \lambda={\Yvcentermath1\Yboxdim{4pt}\yng(2)}\mapsto+
\end{equation}
and
\begin{equation}
    \lambda={\Yvcentermath1\Yboxdim{4pt}\yng(1,1)}\mapsto-,
\end{equation}
the input state is
\begin{align}
\label{eq:input2}
\ket{\bm1}^\pm\left(\bm\tau\right)
 =&\int\frac{\text{d}^2\bm\omega}{\sqrt2}\phi(\bm\omega)\text{e}^{-\text{i}( \omega_1-\omega_2)\tau}  \left(\mathds1\pm P(12)\right)\ket{\bm{1(\omega)}}
   \end{align}
for
\begin{equation}
\bm{\omega}=(\omega_1,\omega_2),\,\bm{\tau}=(\tau,-\tau)^\top.
\end{equation}
For three photons,
six permutations of~$\bm\omega$
are obtained using the permutation group~$S_3$ with elements partitioned into three classes \{$\mathds1$\},
$\sigma_{ab}$, and $\sigma_{abc}$ with
\begin{equation}
 \sigma_{ab}=\{P(12),P(13),P(23)\}   
\end{equation}
and
\begin{equation}
    \sigma_{abc}=\{P(123),P(132)\}
\end{equation}
Replacing
\begin{equation}
    \lambda=\Yvcentermath1\Yboxdim{4pt}\yng(3)\mapsto +
\end{equation} 
and
\begin{equation}
 \lambda=\Yvcentermath1\Yboxdim{4pt}\yng(1,1,1)\mapsto-,   
\end{equation}
the input state
\begin{align}
 \label{eq:input3}
  \ket{\bm1}^\pm(\bm\tau)
=\int&\frac{\text{d}^3\bm\omega}{\sqrt6}
\phi(\bm\omega)\text{e}^{-\text{i}\left(\omega_1-\omega_3\right)\tau}\nonumber\\
\times\bigg[&\mathds1
\pm\left(P(12)+P(13)+P(23)\right)\nonumber\\
&+\left(P(123)+P(321)\right)\bigg]
\ket{\bm1(\bm\omega)},
   \end{align}
with $\bm{\tau}=(\tau,0,-\tau$),
is fully symmetric or alternating, and thus used to obtain the permanent and determinant respectively.
To obtain immanants for $\lambda=$ $\Yvcentermath1\Yboxdim{4pt}\yng(2,1)$ with characters $\{2,0,-1\}$ for the three classes $\mathds1$,
$ \sigma_{ab}$ and $\sigma_{abc}$, the input state is
\begin{align}
 \label{eq:inputimm3}
  \ket{\bm1}^{\Yboxdim{4pt}\yng(2,1)}(\bm\tau)
  	=&\int\text{d}^3\bm\omega\phi(\bm\omega)\text{e}^{-\text{i} \left(\omega_1-\omega_3\right)\tau}
	\nonumber\\
	&\times\frac{1}{\sqrt6}\left[2\mathds1
  -(P(123)+P(321))\right]\ket{\bm1(\bm\omega)}.
   \end{align}
For $n=4$, there are 24 permutation operations with five classes \{$\mathds1$\}, $\sigma_{ab}$,  $\sigma_{(ab)(cd)}$, $\sigma_{(abc)}$ and $\sigma_{(abcd)}$ corresponding to the five partitions.
The input state
has the general form
\begin{align}
 \label{eq:input4}
  &\ket{\bm1}^\lambda\left(\tau\right)
 =\int\frac{\text{d}^4\bm\omega}{{\sqrt{4!}}} \phi(\bm\omega)\text{e}^{-\text{i}\bm\omega\cdot\bm\tau}
  \sum_{\sigma\in S_4}\left[\chi^\lambda(\sigma) \ket{\bm1(\bm\omega_\sigma)}\right],
     \end{align}
where the notation~(\ref{eq:permnotation}) is used and
\begin{equation}
    \bm\tau=(3\tau,\tau,-\tau,-3\tau)^\top.    
\end{equation}
Here
\begin{equation}
  \chi^{\Yvcentermath1\Yboxdim{4pt} \yng(4)}(\sigma)\equiv 1\forall \sigma,\;
	\chi^{\Yvcentermath1\Yboxdim{4pt} \yng(1,1,1,1)}\equiv\pm 1,    
\end{equation}
i.e., even/odd, for the permanent and determinant,
 respectively. The characters needed for the three immanants $\lambda=\Yvcentermath1\Yboxdim{4pt}\yng(3,1)$, $\Yvcentermath1\Yboxdim{4pt}\yng(2,2)$ and $\Yvcentermath1\Yboxdim{4pt}\yng(2,1,1)$  for the five classes are \{3,-1,-1,0,1\}, \{2,0,2,-1,0\} and \{3,1,-1,0,-1\}.

The interferometer maps the input state (\ref{eq:input1}) to an output $U_m\ket{\bm1}^\lambda\left(\bm\tau\right)$. The coincidence probability that all detectors at the output click, after the transformation, 
provides an estimate of the quadratic function
\begin{align}
 \label{eq:wplambda}
  \wp^\lambda\left(\bm\tau\right)= { }^\lambda\langle \bm1|U_m^\dagger\Pi_n U_m|\bm1\rangle^\lambda .
\end{align}
The projection operator
\begin{align}
    \Pi_n:=\prod_{j=1}^{n}\int \text d\Omega_j \ket{1(\Omega_j)}\bra{1(\Omega_j)}
\end{align}
is an $n$-fold product
of single photon projections, each modeling the count in detector $j$ of a single photon, with flat
detection response independent of the frequency $\Omega_j$ of the photon.
This samples the coincidence clicks on all the $n$ output ports;
i.e.,
the coincidence probability is obtained by projecting the output state of the interferometer onto exactly one photon per output port with the detector.

As $\vert\bm1\rangle^\lambda$ is a sum involving weighted
permutations,
the function
\begin{align}
\label{eq:quadfunc}
\wp^\lambda\left(\bm\tau\right)
=&\sum_{\sigma_{1},\sigma_2,\sigma'_{1},\sigma'_2\in S_n} \prod_{\imath,\jmath=1}^n\chi^\lambda(\sigma_1)\chi^\lambda(\sigma_2)U_{\imath,\sigma'_1(\imath)}U^*_{\jmath,\sigma'_2(\jmath)}
\nonumber\\
&\qquad \times a_{\sigma_1\circ\sigma_1',\sigma_2\circ\sigma_2'}(\bm{\tau})
\end{align}
where
\begin{align}
  &a_{\sigma_1\circ\sigma_1',\sigma_2\circ\sigma_2'}(\bm{\tau})\nonumber \\
  &=  \int\frac{\text{d}^n\bm\omega}{n!}
\left|\phi(\bm\omega)\right|^2
\text{e}^{-\text{i}\left(\bm\omega_{\sigma_1\circ\sigma'_1}
-\bm\omega_{\sigma_2\circ\sigma'_2}\right)\cdot\bm\tau},
\label{eq:aasintegral}
\end{align}does not contain \emph{only} the immanant for~$\lambda$ of~$U_n$ but also immanants for~$\lambda$ of matrices~$(U_n)_\sigma$, differing from~$U_n$ by permutations of rows.
The subscripts $\sigma\circ\sigma'$ in the
frequency vectors $\bm{\omega}$ refer to a composition of
permutations. For instance, if $\sigma_1’=(12)$ and $\sigma_1=(132)$, then 
$\sigma_1’ \circ \sigma_1=(13)$.
The output of the algorithm is the estimate of the quadratic function~(\ref{eq:quadfunc}), which we explain in the next subsection.        
\subsection{Estimate of the quadratic function}
\label{subsec:estimate}
To estimate the quadratic function (\ref{eq:quadfunc}), we fix~$n$,
the number of input photons,
the time delays $\bm\tau$ between the photons in each input and the partition~$\lambda$
of~$n$.
Given the input, we have the list of permutations of~$S_n$, a lookup table for the characters $\chi^{\lambda}$ for each partition~$\lambda$, the matrix~$U_n$ and its row-permuted versions.
With these inputs, the sum of the products of the integrals~(\ref{eq:quadfunc}) fully determines what the coincidence rate and hence the quadratic function is.
The integral~(\ref{eq:quadfunc}) is the Fourier transform of the power spectrum and $U_{\imath,\sigma'_1(\imath)}$ relabels the output channel in which the photon is detected. 

For simplicity,
we choose a Gaussian spectral function for the sources.
For a source pulse
with carrier frequency~$\omega_0$ and bandwidth~$\sigma_0$, 
\begin{align}
\phi(\omega)
\equiv&\,G\left(\omega;\omega_0,\sigma_0\right)
:=\frac{\exp\left[-\frac{\left(\omega-\omega_0\right)^2}{4\sigma_0^2}\right]}
{\left(2\pi\sigma_0^2\right)^{1/4}}
\end{align}
and
\begin{equation}
G(\bm\omega;\omega_0,\sigma_0)
:=G(\omega_1;\omega_0,\sigma_0)\times \cdots \times 
G(\omega_n;\omega_0,\sigma_0).    
\end{equation}
Successive time delays are multiples of $\tau$ and equally distributed about zero such that
the $\imath^\text{th}$ component of the vector $\bm{\tau}$ is
\begin{align}
    \tau_\imath=
    \begin{cases}
    (n+1-2\imath)\tau,\;&\text{for } n\text{ even},\\
    \frac{1}{2}(n+1-2\imath)\tau,\;&\text{for } n\text{ odd},
    \end{cases}
    \label{eq:delays}
\end{align}
which leads to each integral~(\ref{eq:aasintegral}) being a function of
\begin{equation}
	\widetilde{G}(\tau,\sigma_0):=\exp(-\sigma^2_0\tau^2).  
\end{equation}
Without loss of generality, $\sigma_0\equiv 1$ so $\tau$ is dimensionless
scaled time,
which quantifies how long the delay should
be in terms of the width of the pulse.
For any general $n$,
we can rewrite the somewhat complicated form of Eq.~(\ref{eq:quadfunc}) in terms of three basic quantities.
 
The first quantity is~$\bm{\Lambda}(\lambda)$,
which is a column vector containing functions of products of immanants of the submatrix.
Finally,
we have~$\bm{A}^\lambda$, whose elements are integers
and the $(K+1)$-dimensional column vector $\bm{\widetilde G}(\tau)$ with component $k$ given by the $(k-1)^\text{th}$ power of 
$\tilde G(\tau)$,
namely,
\begin{equation}
\bm{\widetilde G}(\tau):=
\left((\widetilde{G}(\tau))^0, (\widetilde{G}(\tau))^1, \ldots, (\widetilde{G}(\tau))^K\right)^\top.
\end{equation}

For each $\lambda$, the estimate~(\ref{eq:quadfunc})
can then be written in the cleaner form
\begin{equation}
\label{eq:rate}
\wp^{\lambda}(\tau)
=\bm{\Lambda}^\top(\lambda)
\bm{A}^\lambda\widetilde{\bm G}(\tau),
\end{equation}
where, for the choice of delay times~(\ref{eq:delays}), we have
\begin{align}
K=
    \begin{cases}
    \sum\limits_{\imath=1}^n 2(n+1-2\imath)^2=\frac{2}{3}n(n^2-1),&\text{even } n,\\
     \sum\limits_{\imath=1}^n(n+1-2\imath)^2/2=\frac{1}{6}n(n^2-1),&\text{odd } n,
    \end{cases}
    \label{eq:K}
\end{align}
which fixes the maximum degree of $\widetilde G(\tau)$. 

 The estimate of the quadratic function~(\ref{eq:wplambda}) is achieved as a sampling problem. Specifically, the server repeatedly injects time-bin-entangled multiphoton states and records the coincidences for detectors placed at each output port. The server repeats the experiment $L$ times, and a number $l$ of coincidence events at the output is recorded.
 Using the familiar tools of binomial statistics,
 the event probability
 is estimated to be~\cite{McQuarrie03}
\begin{align}
\frac{l}{L} \pm \frac{z}{L}\sqrt{\frac{l(L - l)}{L}}
\end{align}
with $z$ related to the confidence level.
For a $95\%$ confidence interval, $z \approx 1.96$,
but other confidence levels are possible.
One can increase $L$
to narrow the error on the rate so it is below the desired threshold $\epsilon$ and thus the estimate of the quadratic
function $F_\lambda(U_n)$.

\subsection{Experimental realization}
\label{subsec:exprealize}
We now explain the construction of the interferometer based on the unitary matrix~$U$.
The matrix is decomposed by exploiting the fact that any irrep of $U_m$ can be parametrized in a basis that reduces a particular $m-1$ subgroup 
and it is therefore advantageous to recursively factorize each $SU(m)$ transformation into a product of $SU(2)_{ij}$ subgroup transformations
mixing fields~$i$ and $j$~\cite{dGSBZ01,RSdG99}.
Experimentally, any $U_m$ matrix can be visualized
as a series of $\binom{m}{2}$ beam splitters and phase shifters~\cite{CHM+16,DG15,dGMS18} (see also \cite{urias2010householder}). 
The matrix~$U_n$ is a submatrix of~$U_m$, determined by keeping 
the $n\le m$ rows
of~$U_m$ corresponding to the input channels, and 
the~$n$ columns of~$U$ corresponding to detection channels.
The rows and columns need not be the same, and the $n\times n$ submatrix
need not be unitary.
 
Two photon time-bin entangled states have been created by a single emitter using a Franson interferometer~\cite{JPK+14}. In this method, an
incoming laser pulse is created into a coherent superposition of early and late photons through an unbalanced interferometer~\cite{Franson89}. Two
photon time-bin-entangled states are then generated by creating a biexciton state of a quantum dot, either by the early or by the late
pulse,
followed by the emission of a biexciton–exciton photon cascade~\cite{JPK+14}.
The emitted photons are thus in the time-bin entangled state.
Generating time-bin entangled states beyond the two-photon case is a challenge,
which we leave for future work. 
\section{Quadratic Form for any immanant of $U_n$ for $n=2$, $3$ and $4$}
\label{sec:n234}
\begin{figure}
\begin{minipage}[c]{0.45\linewidth}
\subfloat
[$\lambda={\protect\yongtwo}$]{\label{fig:subA2}\makebox[1.5\width][c]{
	   \centering
	  \includegraphics[scale=1.25]{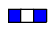}	 
}  
}
\end{minipage}
\begin{minipage}[c]{0.45\linewidth}
\subfloat[b][$\lambda={\protect\yongoneone}$]{\label{fig:subA11}
	   \centering
	  \includegraphics[scale=1.25]{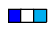}	 
}
\end{minipage}\\
\begin{minipage}[c]{0.45\linewidth}
\subfloat[b][$\lambda={\protect\yongthree}$]{\label{fig:sub3}
	   \centering
	  \includegraphics[scale=1.25]{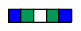}	 
}
\end{minipage}
\begin{minipage}[c]{0.45\linewidth}
\subfloat[c][$\lambda={\protect\yongoneoneone}$
 ]{ \label{fig:subA111}
	   \centering
	 \includegraphics[scale=1.25]{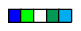}	 
 }
 \end{minipage}
 \\

\begin{minipage}[c]{0.45\linewidth}
  \subfloat[b][$\lambda={\protect\yongtwoone}$
  ]{\label{fig:subA21}
	   \centering
 \includegraphics[scale=1.25]{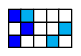}	 

}
\end{minipage}
\begin{minipage}[c]{0.45\linewidth}
\subfloat[c][$\lambda={\protect\yongfour}$
  ]{\label{fig:subA4}
	   \centering
	   \includegraphics[scale=1.25]{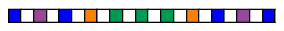}	 
 
}
\end{minipage}\\
\begin{minipage}[c]{0.45\linewidth}
\subfloat[b][$\lambda={\protect\yongfourones}$
 ]{ \label{fig:subA1111}
	   \centering
	   \includegraphics[scale=1.25]{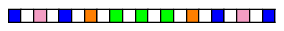}	  
}
\end{minipage}
\begin{minipage}[c]{0.45\linewidth}
\subfloat[b][$\lambda={\protect\yongtwotwo}$
  ]{\label{fig:subA22}
  \centering
	  
	   \includegraphics[scale=0.4]{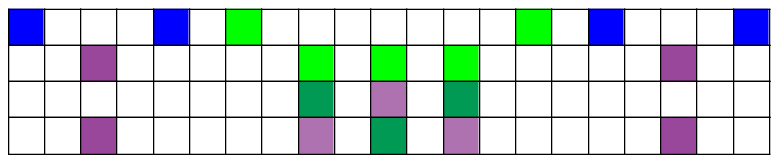}	  
}\end{minipage}\\
\begin{minipage}[c]{0.45\linewidth}
\subfloat[b][$\lambda={\protect\yongthreeone}$
  ]{\label{fig:subA31}
  \centering
\includegraphics[scale=0.4]{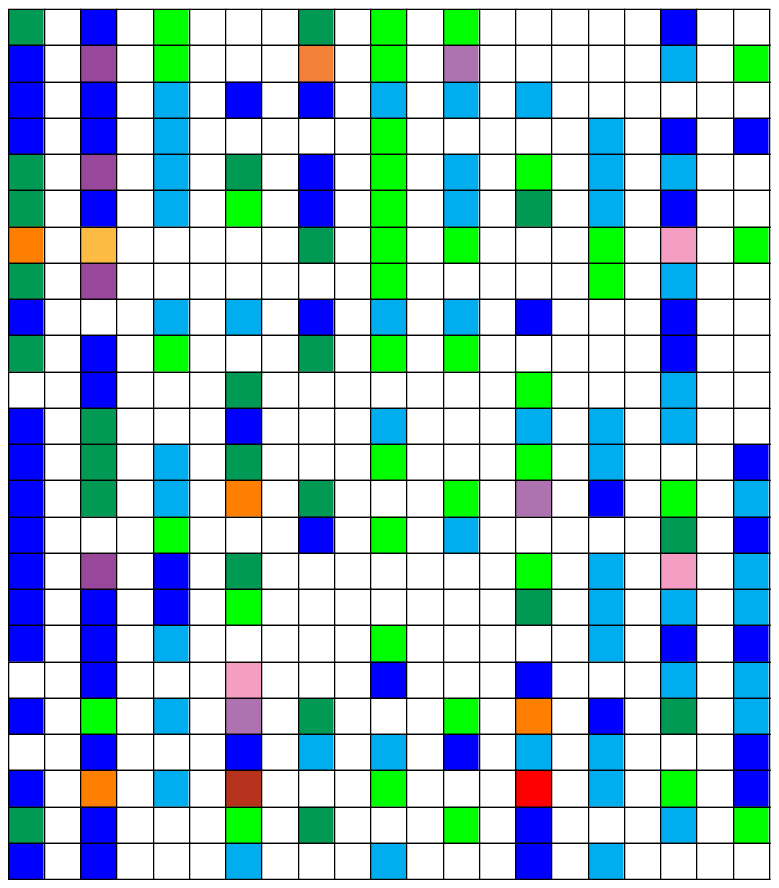}	  	  
}
\end{minipage}
\begin{minipage}[c]{0.45\linewidth}
\subfloat[b][$\lambda={\protect\yongtwooneone}$
]{\label{fig:subA211}
  \centering
\includegraphics[scale=0.4]{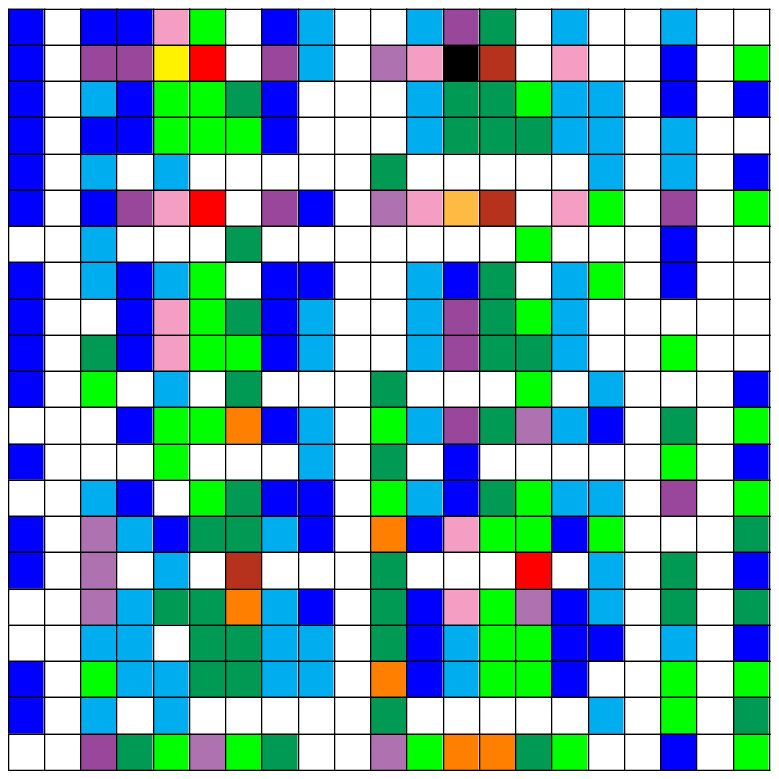}	  
}
\end{minipage}

\caption{$\bm A^\lambda$ for different~$\lambda$. Each element in a row of matrix is the coefficient of various powers of $\widetilde{G}(\tau)$. For even $n$, only coefficients of even powers of $\widetilde{G}(\tau)$ are shown. The colors code the values of coefficient as: $\protect\yongwhite=0$, $\protect\yongCyan=-1$, $\protect\yongblue=1$, $\protect\yonggreen=-2$, $\protect\yongForestGreen=2$, $\protect\yongLavender=-3$, $\protect\yongPurple=3$, $\protect\yongOrchid=-4$, $\protect\yongorange=4$, $\protect\yongyellow=-5$, $\protect\yongDandelion=5$,  $\protect\yongred=-6$, $\protect\yongBrickRed=6$ and $\protect\yongblack=7$.
}
\label{fig:As}
\end{figure}

We now discuss specific realizations of $n\in\{2,3,4\}$ photons scattering inside an $m$-channel  interferometer, with also $n$ detectors to record the arrival of photons. A particular irrep of the permutation group $S_n$ is labeled by a partition conveniently represented by a Young diagram, and so is an immanant; hence
$\lambda$ is used interchangeably for both.
For $n=2$, there are only the permanent and determinant of the $U_2$ matrix. For $n=3$ and $4$,
the number of immanants
is three and five respectively, one for each of the conjugacy classes
of the~$S_3$ or $S_4$ symmetric group, for input states suggested in Eqs.~(\ref{eq:input2}) to~(\ref{eq:input4}). For each $n$ and $\lambda$, we get $\bm A$ from Fig.~\ref{fig:As}, where for even $n$ only coefficients of even powers of $\widetilde G(\tau)$ are shown as the rest are zero.

The number of columns for odd $n$ is $K+1$ and for even $n$ is $K/2+1$. We see that as per~(\ref{eq:K})
\begin{align}
    K=
    \begin{cases}
    4\quad&\text{for } n=2\text{ and }3.\\
    40 \quad&\text{for } n=4.
    \end{cases}
\end{align}
For each $n$ and
\begin{equation}
    \lambda_n^+\mapsto \text{per}(U_n)
\end{equation}
and
\begin{equation}
    \lambda_n^-\mapsto \text{det}(U_n),
\end{equation}
we obtain
\begin{align}
\bm\Lambda({\lambda^{\pm}_n})= \frac{1}{n!} \left|\lambda^\pm_n\right|^2  
\end{align}
and $\bm A^\lambda$ is in Figs.~\subref*{fig:subA2} and \subref*{fig:subA11}, \subref*{fig:sub3} and \subref*{fig:subA111}, and \subref*{fig:subA4} and \subref*{fig:subA1111}, for $n\in\{2,3,4\}$ respectively.

For~$\lambda=\Yvcentermath1\Yboxdim{4pt}\yng(2,1)$, the three-fold coincidence probability estimates the quadratic function~(\ref{eq:rate}) 
with
\begin{align}
\bm{\Lambda}(\lambda)=
\frac{1}{3}
 \begin{pmatrix}
2\left(\left|\lambda\right|^2+\left|\lambda_{23}\right|^2+\left|\lambda_{21}\right|^2 +\left|\lambda_{123}\right|^2\right)\\
\lambda\left( \lambda^*_{21}+2\lambda^*_{23}\right)+\left(\lambda_{23}+\lambda_{12}\right)\lambda^*_{123} \\
\lambda^*\lambda_{123}+\lambda^*_{23}\lambda_{12}
\end{pmatrix},
\end{align}
for $\lambda_{ij}$ and $\lambda_{ijk}$ corresponding to permutations $(ij)$ and $(ijk)$ of rows~$i$, $j$ and $k$ of the matrix~$U_n$, with $(ijk)$ denoting the cycle
$i\to j,j\to k,k\to i$ etc.
Here $\lambda^*$ corresponds to the complex conjugate of~$\lambda$, and $\bm A^\lambda$ is given in Fig.~{\subref*{fig:subA21}}.
The probability~(\ref{eq:rate})
is a quadratic function of the immanant of $U_n$ and its permutations
and linear
in~$\widetilde{\bm G}(\bm\tau)$, for irrep
$\lambda=\Yvcentermath1\Yboxdim{4pt}{\yng(2,1)}$.
The explicit relation between quantities~(\ref{eq:FlambdaU}) and~(\ref{eq:rate}) for $\lambda=\Yvcentermath1\Yboxdim{4pt}\yng(1,1)$ and $\Yvcentermath1\Yboxdim{4pt}\yng(2,1)$ is in Appendix~A.

For $\lambda=\Yvcentermath1\Yboxdim{4pt}\yng(2,2)$,
using input state~(\ref{eq:input4})
with corresponding characters,
the coincidence probability estimates
quadratic function~(\ref{eq:rate})
with
\begin{align}
&\bm\Lambda(\lambda)=
\frac{1}{6}
 \begin{pmatrix}
\xi
\\
 {\lambda}^*_{12}{\lambda}+{\lambda}^*_{124}{\lambda}_{13}\\
{\lambda}^*_{13}{\lambda}\\
{\lambda}^*_{124}{\lambda}_{12}
\end{pmatrix}
\end{align}
for
\begin{equation}
\xi:=\left|{\lambda}\right|^2+\left|{\lambda}_{12}\right|^2+\left|{\lambda}_{13}\right|^2+\left|{\lambda}_{124}\right|^2+{\lambda}_{124}^*{\lambda}
+{\lambda}^*_{12}{\lambda}_{13}    
\end{equation}
and $\bm A^{\lambda}$ is in Fig.~\subref*{fig:subA22}.
The coincidence probability is a quadratic function of one immanant ${\Yvcentermath1\Yboxdim{4pt}\yng(2,2)}$
of a $4\times 4$ submatrix $U_4$ and its permutations.

For $\lambda=\Yvcentermath1\Yboxdim{4pt}\yng(3,1)$ and 
$\Yvcentermath1\Yboxdim{4pt}\yng(2,1,1)$, and sending the input states~(\ref{eq:input4}),
the coincidence probability gives the estimate of the quadratic function $F_\lambda(U)$ 
for each~$\lambda$, with $\bm A^\lambda$ in Figs.~\subref*{fig:subA31} and~\subref*{fig:subA211}, 
respectively, and $\bm\Lambda(\lambda)$ given in Appendix~B.
Except for those partitions corresponding to permanents,
we see from Figs.~\subref*{fig:subA2},~\subref*{fig:sub3} and~\subref*{fig:subA4}
that the sum of 
elements of $\bm A$ in each row equals zero for 
all $n$. Thus,  for indistinguishable photons ($\tau=0$), all coincidence 
probabilities vanish except 
$\wp^{\Yvcentermath1\Yboxdim{4pt}\yng(2)\cdots\Yvcentermath1\Yboxdim{4pt}\yng(1)}$,
which becomes equal to the permanent.

The precise structure of $\bm{\Lambda}(\lambda)$, 
especially as a function of the number of photons in the system,
is unknown but intriguing.
The answer would depend on linear relations between immanants
of shape $\lambda$, and these relations are apparently not known yet.
As an example,
for $\lambda=\Yvcentermath1\Yboxdim{4pt}\yng(2,1)$, the sum of immanants
with columns permuted is $0$; on the other hand no such relation exists for immanants
of the type $\Yvcentermath1\Yboxdim{4pt}\yng(3,1)$.  
Moreover, this sum rule is certainly not true of sums of permanents
of matrices with permuted columns since the permanent is invariant under such 
permutations.    

As a result, the expressions for $\bm{\Lambda}(\lambda)$ are not unique, 
and there
is currently no obvious guide in ascertaining which set of linearly independent
immanants will provide improved insight into these expressions.
In one case expressions for some rates were simplified
by choosing a set of immanants that were not all linearly independent~\cite{dGTP+14}.
Structural issues of $\bm{\Lambda}(\lambda)$ remain at this stage
open questions.

\section{Conclusions}
\label{sec:conclusions}
In conclusion,
we propose a scheme
that estimates a quadratic function of a specific immanant of a unitary matrix~$U_n$,
which describes 
an $m$-channel $n$-photon interferometric transformation,
and permutations of rows of~$U_n$.
Our results gives actual quantum physical meaning to the immanant in the sense that the multiphoton output coincidence rate samples a quadratic function $F_\lambda(U)$, while photon
coincidences in previous studies depended on all of the immanants including the permanent and determinant so showed dependence on immanants but did not isolate one immanant as we have done here.

In order to translate our concept to experimental realization,
we explained how to decompose~$U_m$
into a configuration of beam splitters and phase shifters.
In contrast to previous studies,
which consider only multiphoton input states that are product,
here we have introduced time-bin-entangled input states,
which still feature zero or one photon entering each input port but arrival times are entangled.
Constructing the input state is a challenge,
although a post-selected variant of time-bin-entangled states could be achievable as an extension of Franson's post-selected superposition of two photons with one being early and the other late with the case that the first photon is late and the second one early~\cite{Franson89}.
Devising multiphoton time-bin entangled states is challenging,
which we save for future study. For two-photon coincidences, recent experiments report around $10^4$ coincidence counts per second~\cite{BGD+13}. Though experiments are not being performed yet for more photon coincidences, these figures suggest a reasonable count rate for that as well.

The fact that input states are suggested to be generated by postselection methods will not affect the efficiency of the scheme compared to those based on single photons. Postselected input states such as parametric down conversion states have been suggested in scattershot BosonSampling~\cite{LLR-K+14} and experiments show that the event rate is 4.5 times better than for BosonSampling experiments with single photons~\cite{BSV+15},
 due to inefficient generation of photons in the latter~\cite{EFM+11}. We expect that probabilistic sources in our case will not introduce an added disadvantage over sampling based on single-photon sources.
 
The advantage of our scheme over classical algorithms needs to be analyzed.
 The calculation of the permanent using sampling does not provide an exponential speed-up, compared to the classical algorithms~\cite{LBR17,AA11,MN15,JSV04}, and we expect the same for immanants.
 In addition, losses and imperfections further degrade the efficiency of this scheme. The analysis of BosonSampling with photon losses, dark counts and other imperfections~\cite{AB16,R-KRC16} shows that a quantum system is classically simulatable if loss is constant. In fact any quantum circuit with constant loss is classically simulatable~\cite{GD18}. 
However for real experiments, the quantum advantage is not based on the computational complexity and the physical requirement of the quantum speedup but on the difficulty of simulating these experiments classically~\cite{LLR-K+14,GD18}.

 We have constructed the theory generally for any~$n$ and applied it to $n\in\{2,3,4\}$. The quadratic functions of each immanant are sorted out by time-bin entangled state,
which gives the immanant an interferometric realization. Ideally input states could be engineered to make only the coefficient of one immanant of $U_n$ to survive and that of its permutations to go to zero. This would require a different experimental setup and input state, which can be explored as future research.

\acknowledgments
We thank Dr. A.\ Khalid for providing computer code to generate immanants of of a $4\times 4$ matrix
and we appreciate financial support from NSERC and from the American Physical Society through their International Research Travel Award Program.

\begin{widetext}
\section*{Appendix}
 \subsection{Relating Eqs.~(\ref{eq:FlambdaU}) and (\ref{eq:rate})}
 \label{app:relatauG}
The summation of Eq.~(5) can be expressed more compactly using the matrix formulation of (37).  This requires the explicit computation of the coefficients $a_{\sigma\sigma'}$, some of which will be identical for different $(\sigma\sigma')$ pairs.  A direct calculation shows that each $a_{\sigma\sigma'}$ is proportional
to one entry 
of the matrix product $\bm A^{\lambda} \widetilde{\bm G}(\tau)$. 

For $\lambda=\Yvcentermath1\Yboxdim{4pt}\yng(1,1)$, we get
\begin{align}
 a_{(I)(I)} &=\frac{1}{2} \bm{A}^\lambda\widetilde{\bm G}(\tau\\
 &=\frac{1}{2}\left(1-4\text e^{-4\sigma_0^2\tau^2}\right)
\end{align}
 
For $\lambda=\Yvcentermath1\Yboxdim{4pt}\yng(2,1)$, four linearly independent immanants
and ten coefficients $\{a_{\sigma\sigma'}\}$ exist.
From our Fig.~2,
we read
\begin{align}
\bm A^{\lambda}=
    \begin{pmatrix}
    1&-1&0&0&0\\
    0&1&0&0&-1\\
    1&0&0&-1&0
    \end{pmatrix}
\end{align}
and, from Eq.~(36), we have
\begin{align}
    \widetilde{\bm G}(\tau)=
    \begin{pmatrix}
    1\\
    \text e^{-\sigma_0^2\tau^2}\\
    \text e^{-2\sigma_0^2\tau^2}\\
    \text e^{-3\sigma_0^2\tau^2}\\
    \text e^{-4\sigma_0^2\tau^2}
    \end{pmatrix}.
\end{align}
A direct comparison between the entries of $\bm A^{\lambda} \widetilde{\bm G}(\tau)$ and the integral expression of $a_{\sigma\sigma'}=a_{\sigma'\sigma}$ in (\ref{eq:FlambdaU}) yields
\begin{align}
    a_{(I)(I)}=a_{(23)(23)}=a_{(12)(12)}=a_{(123)(123)}&=\frac{2}{3}\sum_{\imath}\bm A^\lambda_{1\imath} \widetilde{\bm G}_{\imath 1}(\tau)=\frac{2}{3}\left(1-e^{\sigma_0^2\tau^2}\right)\\
    a_{(I)(12)}=a_{(23)(123)}=a_{(12)(123)}&=\frac{1}{3}\sum_{\imath}\bm A^\lambda_{2\imath} \widetilde{\bm G}_{\imath 2}(\tau)=-\frac{1}{3}\left(e^{-4\sigma_0\tau^2}+e^{-\sigma_0^2\tau^2}\right) \\
    a_{(I)(23)}&=\frac{2}{3}\sum_{\imath}\bm A^\lambda_{2\imath} \widetilde{\bm G}_{\imath 2}(\tau)
    =\frac{2}{3}\left(e^{-4\sigma_0\tau^2}+e^{-\sigma_0^2\tau^2}\right) \\
    a_{(I)(123)}=a_{(23)(12)}&=\frac{1}{3}\sum_{\imath}\bm A^\lambda_{3\imath} \widetilde{\bm G}_{\imath 3}(\tau)=\frac{1}{3}\left(1-e^{-3\sigma_0^2\tau^2}\right).
    \end{align}
The computation of the rate is concluded by multiplying each of these terms by the appropriate entries of the 
$\bm{\Lambda}$ matrix, which contains all the information about the linear combinations of immanants that occur once the various terms of Eq.~(5) have been collected.

\subsection{$\bm\Lambda(\lambda)$ for $n=4$ }
\label{app:n4Lambda}
Single-photon coincidences at $n\le m$ ports of an $m$-channel interferometer leads to the estimate of the quadratic function, for each $\lambda$, for interaction times multiples of $\tau$,
given in Eq.~(\ref{eq:rate}).
Here we detail the case where $n=4$.  The possible Young diagrams are
$\Yvcentermath1\Yboxdim{5pt}\yng(4),\yng(3,1),\yng(2,2),\yng(2,1,1)$ and~$\Yvcentermath1\Yboxdim{5pt}\yng(1,1,1,1)$.
The first and the last correspond to fully symmetric and fully antisymmetric functions
of the columns (or rows) of a matrix:
the permanent and the determinant respectively.

The character table for~$S_4$ is given in the table below.
\begin{table}[h!]
\label{table:character4}
\begin{center}
\caption{Character table for $S_4$.}
\begin{tabular}{l|ccccc}
 $\hbox{\text{class}}\setminus \hbox{\text{irrep}}$&$\Yvcentermath1\Yboxdim{7pt}\yng(4)$&$\Yvcentermath1\Yboxdim{7pt}\yng(3,1)$
&$\Yvcentermath1\Yboxdim{7pt}\yng(2,2)$&$\Yvcentermath1\Yboxdim{7pt}\yng(2,1,1)$
&$\Yvcentermath1\Yboxdim{6pt}\yng(1,1,1,1,1)$\\ 
\hline
$\chi([4])$     &1 &-1&0&1&-1\\
$\chi([13])$    &1 &0&-1&0&1\\
$\chi([2^2])$   &1&-1&2&-1&1\\
$\chi([1^2 2]))$&1&1&0&-1&-1\\
$\chi([1^4])$   &1&3&2&3&1
\end{tabular}
\end{center}
\end{table}

so that, for instance, the $\Yvcentermath1\Yboxdim{6pt}\yng(3,1)$-immanant for the matrix
$
U=\left(\begin{array}{cccc}
U_{11}&U_{12}&U_{13}&U_{14}\\
U_{21}&U_{22}&U_{23}&U_{24}\\
U_{31}&U_{32}&U_{33}&U_{34}\\
U_{41}&U_{42}&U_{43}&U_{44}
\end{array}\right)
$
is
\begin{align}
\text{imm}^{\Yvcentermath1\Yboxdim{5pt}\yng(3,1)}=
&-U_{14} U_{23} U_{32} U_{41}-U_{13} U_{24} U_{32} U_{41}+U_{14} U_{22} U_{33} U_{41}-U_{12} U_{23} U_{34} U_{41}
-U_{14} U_{23} U_{31} U_{42}\nonumber \\
&
-U_{13} U_{24} U_{31} U_{42}
+U_{11} U_{24} U_{33} U_{42}-U_{13} U_{21} U_{34} U_{42}-U_{12} U_{24} U_{31} U_{43}-U_{14} U_{21} U_{32} U_{43}
\nonumber \\
&-U_{12} U_{21} U_{34} U_{43}+U_{11} U_{22} U_{34} U_{43}
+U_{13} U_{22} U_{31} U_{44}
+U_{11}U_{23} U_{32} U_{44}\nonumber \\
&+U_{12} U_{21} U_{33} U_{44}+3 U_{11} U_{22} U_{33} U_{44}.
\end{align}

For $n=4$ and for $\lambda=\Yvcentermath1\Yboxdim{5pt}\yng(3,1)$, our vector $\Lambda(\lambda)$ has the form
 \begin{align}
 \Lambda(\lambda)=
 \frac{1}{24}
     \begin{pmatrix}
       2\left(\left|{\lambda}\right|^2
      -{\lambda}^*{\lambda}_{12}\right)\\
      \left|{\lambda}_{12}\right|^2+\left|{\lambda}_{34}\right|^2+\left|{\lambda}_{124}\right|^2\\
      -2\left(\left(2\,{\lambda}+2\,{\lambda}_{12}-{\lambda}_{34}\right){\lambda}^*_{14}\right)\\
      2\left|{\lambda}_{13}\right|^2
 +2{\lambda}_{13}{\lambda}^*-{\lambda}_{14}{\lambda}^*_{132}\\
 2\left|{\lambda}_{14}\right|^2\\
 2\left(\left|{\lambda}_{23}\right|^2+\left|{\lambda}_{132}\right|^2\right)\\
 \left|{\lambda}_{143}\right|^2\\
 {\lambda}^*_{12}{\lambda}_{34}\\
 -2\left(\left(2\,{\lambda}-{\lambda}_{12}-{\lambda}_{43}\right){\lambda)}^*_{23}+\left({\lambda}_{12}+{\lambda}_{124}\right){\lambda}^*_{132}\right)\\
 -2{\lambda}^*{\lambda}_{34}\\
 2{\lambda}^*{\lambda}_{124}\\
 {\lambda}_{14}{\lambda}^*_{143}\\
 2\left(\left({\lambda}+{\lambda}_{12}\right){\lambda}^*_{143}\right)\\
  -{\lambda}^*_{12}{\lambda}_{124}\\
  2{\lambda}_{14}{\lambda}^*_{23}\\
  -{\lambda}_{14}{\lambda}^*_{124}\\
  {\lambda}_{23}{\lambda}^*_{124}\\
\left(2{\lambda}+{\lambda}_{14}-{\lambda}_{23}\right){\lambda}^*_{132}
+2\left({\lambda}_{12}+{\lambda}_{14}+{\lambda}_{23}
 +{\lambda}_{43}-{\lambda}_{132}-{\lambda}_{143}\right){\lambda}^*_{13}\\
 2{\lambda}_{23}{\lambda}^*_{143}\\
 -{\lambda}_{34}{\lambda}^*_{124}\\
 -2{\lambda}_{34}{\lambda}^*_{132}\\
 -{\lambda}_{34}{\lambda}^*_{143}\\
 {\lambda}_{124}{\lambda}_{143}^*\\
 2{\lambda}_{132}{\lambda}^*_{143}
     \end{pmatrix}
     +\text{cc}
 \end{align}

For the irrep
$\lambda=\Yvcentermath1\Yboxdim{4pt}\yng(2,1,1)$,
we have in turn
 \begin{align}
 \Lambda(\lambda)=\frac{1}{24}
     \begin{pmatrix}
 2\left(\left|{\lambda}\right|^2+  {\lambda}\left({\lambda}^*_{12}+{\lambda}^*_{34}+{\lambda}^*_{143}\right)\right)\\
     \left|{\lambda}_{12}\right|^2+\frac{1}{2}\left(\left|{\lambda}_{34}\right|^2 +
   \left|{\lambda}_{124}\right|^2\right)\\
   \left(\left|{\lambda}_{14}\right|^2\right)\\
   \left(\left|{\lambda}_{23}\right|^2+\left|{\lambda}_{132}\right|^2\right)\\
   2\left(\left|{\lambda}_{13}\right|^2+\left({\lambda}+{\lambda}_{12}+{\lambda}_{14}+{\lambda}_{34}+{\lambda}_{23}+{\lambda}_{132}+{\lambda}_{143}\right){\lambda}^*_{13}+{\lambda}_{132}{\lambda}^*_{14}\right)\\
   \frac{1}{2}\left|{\lambda}_{143}\right|^2\\
   2{\lambda}^*_{124}{\lambda}\\{\lambda}^*_{12}{\lambda}_{34}\\
   \left(2\,{\lambda}^*_{14}\left({\lambda}+{\lambda}_{12}+{\lambda}_{34}+\right)+{\lambda}^*_{34}\left({\lambda}_{23}\right)\right)\\
   {\lambda}^*_{23}\left(2\,{\lambda}+{\lambda}_{12}\right)+{\lambda}^*_{132}\left({\lambda}_{12}+{\lambda}_{124}\right)\\
   {\lambda}^*_{12}{\lambda}_{143}\\
  {\lambda}^*_{12}{\lambda}_{124}\\
  2{\lambda}^*_{14}{\lambda}_{23}\\
  {\lambda}^*_{14}{\lambda}_{124}\\
  {\lambda}^*_{34}{\lambda}_{132}\\
  {\lambda}^*_{34}{\lambda}_{143}\\
  {\lambda}^*_{34}{\lambda}_{124}\\
  {\lambda}^*_{23}{\lambda}_{124}\nonumber\\
  {\lambda}^*_{23}{\lambda}_{143}\\
  {\lambda}^*_{23}{\lambda}_{132}\\{\lambda}^*_{143}{\lambda}_{124}
 \end{pmatrix}
  +\text{cc}
 \end{align}
 which is a quadratic function of immanant $\lambda$ of the $U_n$ matrix and its permutations.
 \end{widetext}
\bibliographystyle{apsrev4-1}
\bibliography{references}
\end{document}